\documentclass[review,authoryear,11pt]{elsarticle}

\usepackage[gen]{eurosym}
\usepackage{footnote}
\usepackage{array}
\usepackage[top=1in,bottom=1in,left=1.25in,right=1.25in]{geometry} 

\usepackage{secdot}
\usepackage[subrefformat=parens,labelformat=parens]{subfig}
\usepackage{latexsym}
\usepackage{graphicx}
\usepackage{caption}
\usepackage{enumitem}
\usepackage{float}
\usepackage{color}
\usepackage{wrapfig}
\usepackage{fancyhdr}
\usepackage[hyphens]{url}
\usepackage{natbib}
\usepackage{apalike}
\usepackage[pagewise]{lineno}

\usepackage{xargs}                      
\usepackage[pdftex,dvipsnames]{xcolor}  

\usepackage[colorinlistoftodos]{todonotes} 

\usepackage{setspace}
\usepackage{color}
\definecolor{EditBlue}{RGB}{0,128,255}
\newcommand{\eb}[1]{{\color{black}#1}}

\journal{European Journal of Operational Research}

\begin{document}

\begin{frontmatter}

\title{Perspectives on How To Conduct Responsible \\ Anti-Human Trafficking Research in Operations and Analytics}


\author[FBS]{Renata A. Konrad\corref{cor1}}
\cortext[cor1]{Corresponding author}
\ead{rkonrad@wpi.edu}

\author[NEU]{Kayse Lee Maass}
\ead{k.maass@northeastern.edu}

\author[WPIDS]{Geri L. Dimas}
\ead{gldimas@wpi.edu}

\author[FBS,WPIDS]{Andrew C. Trapp}
\ead{atrapp@wpi.edu}

\address[FBS]{\eb{School of Business}, Worcester Polytechnic Institute, 100 Institute Road, Worcester, MA 01609 USA}
\address[NEU]{ \eb{Department of Mechanical and} Industrial Engineering, \eb{Northeastern University}, 360 Huntington Ave, Boston, MA 02115 USA}
\address[WPIDS]{Data Science Program, Worcester Polytechnic Institute, 100 Institute Road, Worcester, MA 01609 USA}


\begin{abstract}
Human trafficking, the commercial exploitation of individuals, is a gross violation of human rights; harming societies, economies, health and development. The related disciplines of  Operations Research (OR) and Analytics are uniquely positioned to support trafficking prevention and intervention efforts by efficiently evaluating a plethora of decision alternatives and providing quantitative, actionable insights. As operations and analytical efforts in the counter-trafficking field emerge, it is imperative to grasp subtle, yet distinctive, nuances associated with human trafficking. This \eb{paper} is intended to inform those practitioners working in the Operations and Analytics fields by highlighting key features of human trafficking activity. We grouped \eb{ten} themes around two broad categories: (1) representation of human trafficking and (2) consideration of survivors and communities. These insights are derived from our collective experience in working in this area and substantiated by domain expertise. Based on these areas, we then suggest avenues for future work.

\end{abstract}


\end{frontmatter}

\section{Definition Of Human Trafficking} \label{Sec_Intro}


\looseness-1 Human trafficking (HT) is more prevalent and global than many imagine. The United States considers ``trafficking in persons'' and ``human trafficking'' to be interchangeable umbrella terms that refer to both sex trafficking and labor trafficking~\citep{us22}.  It is the crime of using force, fraud, or coercion for the purpose of compelled labor or  commercial sex~\citep{usd18}. Despite the name, trafficking doesn't necessitate the movement of victims but rather an individual's inability to escape exploitation due to reasons such as debt bondage, threats to themselves or family, shame, and fear of deportation (in cases of transnational exploitation), among others.

\looseness-1 Broadly speaking, human trafficking is categorized as (i) sex trafficking -- which includes trafficked labor within escort services, illicit massage business, brothels, and pornography, and (ii) labor trafficking -- which is commonly found within agricultural, domestic work, food service, and construction, but has also been found in nearly every industry sector. An individual can be a victim of both labor and sex trafficking.  Sex and labor trafficking occur in every country, and  are criminalized in most~\citep{usd18}. Importantly, legal definitions of human trafficking differ from country to country, and even jurisdictions within a country~\citep{far20}. In addition, the operational definition of trafficking (i.e., how it is enforced and reported) greatly differs between countries. Both of these features  contribute to the difficulty of obtaining accurate prevalence estimates.

\looseness-1 Although illicit, human trafficking is often ``hidden in plain sight'' - trafficking co-exists in legitimate supply chains, economic markets and industries. Human trafficking activity also occurs in nefarious environments which the general public may never see. Insights into the structure and characteristics of human trafficking networks are limited, primarily originating from social scientists who have studied human trafficking networks and supply chains from a broad typology perspective~\citep{mar14, mar17, dan14, owe15}. The illicit and criminal nature of trafficking hinders measuring the magnitude of human trafficking activity globally; reporting data is incomplete and cases are severely underreported. Most estimates indicate that tens of millions of adults and children of all gender identities are victimized, representing a multi-billion-dollar global industry~\citep{ilo17}.

\looseness-1  The related disciplines of Operations Research (OR) and Analytics are uniquely positioned to help disrupt this illicit sector by their ability to represent complex systems, efficiently evaluate a plethora of decision alternatives, and provide quantitative, actionable insights into the resulting effects of interventions. Specific aspects of anti-human trafficking disruption efforts such as societal or spatiotemporal characteristics are rarely considered, let alone expressed and incorporated into OR models, with few exceptions, e.g.~\cite{maa20}. While still an emerging area, applications of OR and Analytics to HT have been steadily increasing over the past four years (e.g.~\cite{cau19, cun18,kon19, kon17, maa20, boe19,rab18,nag20}; see Section \ref{Sec_Lit} for a more detailed discussion).

\looseness-1  It is imperative to characterize the perspectives of the myriad stakeholders involved in disrupting human-trafficking networks. While law enforcement may wish to wait to bust a trafficking operation until more evidence can be gathered, possibly at the expense of further trauma to victims, an NGO may wish to \eb{help} the victim(s) \eb{leave} an exploitative situation as soon as possible. Traffickers are likely consumed with minimizing detection or maximizing power, profit or, in OR parlance, ``throughput of illicit goods (here, victims)'', whereas the welfare of victims is often a common goal of all but the traffickers and exploiters. The perspective of victims and survivors of human trafficking need to be incorporated and prioritized in intervention efforts.

\looseness-1 An initial reaction of some OR practitioners may be to apply existing methods to combat trafficking -- for example network interdiction. While the fundamental ideas of such methods are applicable to human trafficking, many nuances of human trafficking are not captured in current methods. For example, unlike illicit consumables such as drugs or ammunition, a person can be exploited repeatedly~\citep{log09,jay13}. Furthermore, whereas drugs, nuclear material, or weapons can easily be tested to determine if an illicit product is present, ascertaining whether a person is a victim of human trafficking is not as straightforward. Trafficking victims may not \eb{recognize they are a victim of trafficking or want to use that label}, leading to further challenges when pursuing prosecution for traffickers. In~\cite{kon17}, both challenges and opportunities for OR and Analytics practitioners working in this area are outlined.

\looseness-1 As OR and Analytics practitioners working in this area, we advocate that for existing OR and Analytics methodologies to successfully disrupt human trafficking networks, the OR and Analytics communities need to grasp subtle yet distinctive nuances associated with human trafficking and to identify appropriate modeling components to properly measure and evaluate interventions within illicit networks. Such efforts need to be guided by domain expertise.

\looseness-1 Our objective is to inform modeling efforts designed to disrupt human trafficking activities by identifying key features of human trafficking activity. We grouped \eb{ten} themes around two broad categories: (i) representation of human trafficking, and (ii) consideration of survivors and communities. These insights are derived from our collective experience in working in this area and reviews of the literature, while being substantiated by domain expertise. Our intent is to emphasize aspects of human trafficking that need to be considered in any responsible OR research if it is to be successful in disrupting this nefarious activity\eb{, as well as} suggest avenues for further research.

\looseness-1 Section~\ref{Sec_Lit} provides the reader with a survey of the recent landscape of OR and Analytics research applied to human trafficking, that in part motivates how human trafficking can be appropriately represented in OR and Analytics studies as more thoroughly discussed in Section~\ref{Sec_Representation_HT}. Section~\ref{Sec_Communities} addresses how OR and Analytics can be used to empower survivors and communities. Bearing in mind how to incorporate the nuances found in human trafficking activity, Section~\ref{Sec_Future_Research} suggests research directions for the OR and Analytics communities, while Section~\ref{Sec_Conclusion} concludes the paper. 
\section{Landscape Of Operations And Analytics Research In Human Trafficking}\label{Sec_Lit}

\looseness-1 Anti-trafficking initiatives would do well to customize approaches to the varied context of human trafficking, instead of simply borrowing best practices from the military or commercial sectors. While interdisciplinary and intersectoral collaborations will have to be considerably intensified for anti-trafficking operations to be more efficient and effective, \eb{strengthening collaborations is not enough. It is also necessary to develop innovative technologies, policies, and economic mechanisms}. Thus, research has a critical role to play in helping practitioners find solutions to the challenges associated with anti-trafficking operations. Understanding the current landscape of works in OR and Analytics can assist in this effort by demonstrating areas of growth for anti-trafficking efforts in these disciplines. 

\begin{figure}[!h]
\includegraphics[width=\textwidth]{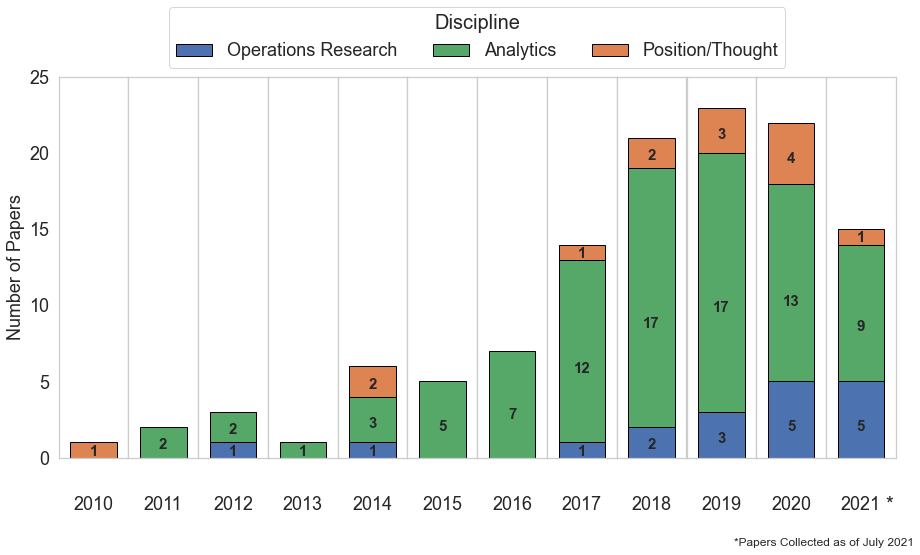}
\caption{Anti-HT studies categorized by OR, Analytics, and Position/Thought~\citep{dim21}.}
\label{fig:Paper_Year_Category}
\end{figure}

\begin{figure}[!h]
\includegraphics[width=\textwidth]{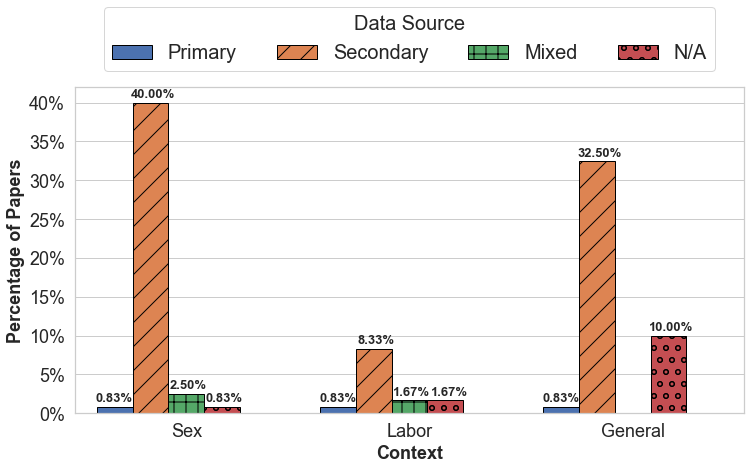}
\caption{ Type of sourced data in OR and Analytics related anti-HT papers. Reprinted
 from \cite{dim21}. Reprinted with permission.
}
\label{fig:Paper_Context_DataType}
\end{figure}

\looseness-1  Dimas et al. (2021) conducted a broad survey of studies that explicitly deal with applications of OR and Analytics to anti-human trafficking efforts over the last 10 years. Publications were categorized into OR, Analytics, and Position/Thought \eb{(such as a perspective piece, providing a framework or research agenda)}, and further distinguished methodological versus position papers. Of the publications reviewed, \eb{73\%} were Analytics-focused, \eb{15\%} of papers were Operations Research-focused and \eb{12\%}  were categorized as Position/Thought pieces~\citep{dim21}. From that review, we present Figures~\ref{fig:Paper_Year_Category} and~\ref{fig:Paper_Context_DataType} which illustrate several trends observed in anti-HT OR and Analytics studies. 

\looseness-1 Figure~\ref{fig:Paper_Year_Category} depicts the total number of studies across the three categories, which while encouragingly increasing over the past decade, still reveals a nascent field. While Analytics-focused studies constitute the largest portion of publications, the number of OR studies is steadily increasing. Although the growth in OR and Analytics studies associated with human trafficking shows promise, there exists gaps in the literature.

\looseness-1 A commendable trend  of the use of real data, be it primary or secondary, is also observed in the studies considered in~\cite{dim21}. Primary data are sourced directly from anti-trafficking organizations or expert interviews / surveys, \eb{whereas} secondary data come from publicly available sources such as census data, websites (e.g., backpage.com and  rubmaps.ch), and government reports. Figure~\ref{fig:Paper_Context_DataType} shows the breakdown of data sources by type of trafficking (sex, labor, or \eb{general applications}), with `mixed' indicating data from both primary and secondary sources. We note that secondary data are the most common source, and particularly so for sex trafficking studies. This is unsurprising, as many researchers are limited to available data such as escort website data from sources similar to the \eb{now} defunct backpage.com. For example, many Analytics-based studies share a common goal of classifying escort ads indicative of sex trafficking. While we recognize the novelty in different approaches to classify ads, the lack of accessible HT data has led to an unbalanced view of trafficking supported by OR and Analytics approaches \eb{that heavily focus} on sex trafficking (further discussed in Section \ref{Sec_Theme_OverRep}). From the review of the literature, two clear avenues emerge for further research: 1) the need for collecting more diverse data for both sex and labor trafficking and 2) the need for the application of OR / Analytics methods specific to labor trafficking.

\looseness-1 ~\citet{dim21} highlight how human trafficking has been studied by the OR and Analytics communities. Research methodologies are explored and deeper insights are provided into the current research gaps. Without understating the depth and breadth of impact made by studies covered in~\cite{dim21}, we emphasize that proper understanding of the nuances of the human trafficking context is imperative for OR and Analytics academics, lest they risk missing the mark by simply adapting methodology meant for military or commercial operations to anti-HT contexts. The social and economic factors which create conditions to make individuals susceptible to exploitation and trafficking, and the complex nature of trafficking operations, need to be considered and incorporated for OR and Analytics to have a positive and sustained impact on anti-HT research. In what follows, we outline \eb{ten} themes to guide responsible OR and Analytics research in the context of human trafficking. 

\section{Representation Of Human Trafficking}\label{Sec_Representation_HT}

We next reflect on several themes that shed light on how human trafficking can be appropriately represented in OR and Analytics studies.

\subsection{Theme 1: Terminology Matters}

\looseness-1 While definitions of what constitutes human trafficking differ between jurisdictions and among human trafficking researchers, the International Labour Organization (ILO)~\citep{ilo12} and the United Nations Palermo Protocol~\citep{un00} provide standardized, widely accepted definitions. Once a state ratifies and implements this legal instrument, country-specific legal frameworks provide comprehensive guidance on the set of criteria for classifying a case as labor or sex trafficking, depending on the criminal or civil laws that have been violated.

\begin{figure}[!h]
\includegraphics[width=\textwidth,height=0.4\textheight]{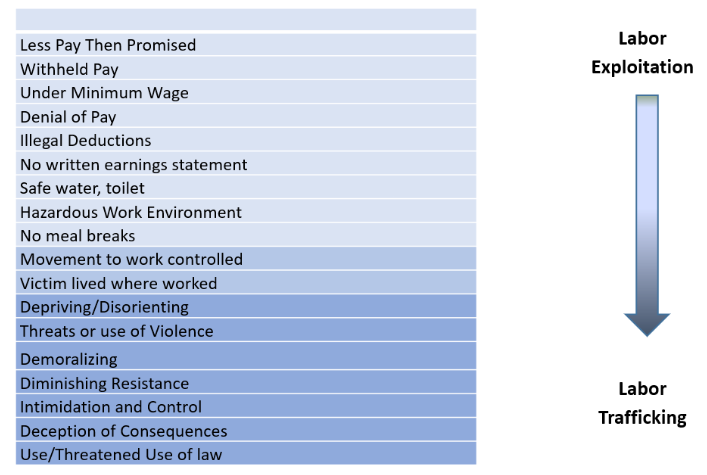}
\caption{Spectrum of Labor Exploitation and Labor Trafficking~\citep{owe15}.}
\label{fig:Labor_trafficking _spectrum}
\end{figure}

\looseness-1 As with many fields in which OR and Analytics applications are relatively new, understanding the terminological landscape of human trafficking establishes credibility and enables understanding and differentiation of the complexities of illicit versus licit markets and networks. For instance, smuggling is not synonymous with trafficking, as a person may voluntarily pay a smuggler to guide them across a border; not all commercial sex work involves sex trafficking and some commercial sex work involves labor trafficking; and exploitation can be mapped onto a spectrum as depicted in Figure~\ref{fig:Labor_trafficking _spectrum},  ranging from what the ILO refers to as ``decent work''~\citep{ilo16} at one end,through various labor and criminal law violations, to extreme exploitation or ``forced labor''. 

\looseness-1 Importantly, ``victim'' and ``survivor'' are commonly used to refer to individuals who have been trafficked. Both terms are important and have different implications when used in the context of victim advocacy and service provision~\citep{ovc11}. Whereas ``victim'' has legal implications within the justice process and refers to an individual who suffered harm as a result of criminal conduct or civil violations,  ``survivor'' is widely used in service-providing organizations to acknowledge the resiliency of people who have experienced victimization~\citep{ovc11}.  

\looseness-1 It is important for OR and Analytics practitioners to be aware of these and similar terminologies given their many implications in designing analytical methods that minimize bias and harm, estimating prevalence measures, and providing access to legal protections.

\subsection{Theme 2: Overrepresentation and Sensationalism}\label{Sec_Theme_OverRep}

\looseness-1 Similar to terminology, it is important to be knowledgeable about the broad landscape of trafficking activity; there is an unfortunate tendency to (i) primarily frame, at least in the United States, human trafficking as a criminal justice issue \eb{for which prosecuting traffickers is the primary metric rather than incorporating a human rights or public health lens} and (ii) conflate human trafficking with sex trafficking. Years of working alongside trafficking researchers and survivors have shown the influence \eb{media overrepresentation of sex trafficking and sensationalism can have on} OR modeling efforts. Inaccurate problem framing will lead to less than useful solutions and may very well perpetuate harm.

\looseness-1 For example, in the United States depictions of human trafficking in movies, documentaries, and television episodes tend to follow a rescue narrative, where innocent victims are saved from harmful predators~\citep{aus17,alb17}. Traffickers are routinely portrayed as part of larger organized crime rings, despite empirical evidence that often points to the contrary~\citep{aus17}. Incorrect framing of human trafficking in the popular media may lead policymakers and legislators to adopt misdirected anti-trafficking responses, particularly responses focused on criminal justice system solutions~\citep{aus17}. While modern human trafficking was first examined through the lens of human rights, the passage in 2000 of the Victims of Trafficking and Violence Protection Act (TVPA) as a Federal US statute helped ``usher in the criminal justice frames in the media''~\citep{far09}. Dominant media themes in the United States continue to cast trafficking in the light of criminal activity, rather than as a human rights, policy, public health, or activist awareness issue~\citep{far09,gul10,joh14,cha15,joh15}; one study suggests that only approximately 30\% of sampled print media attempted to relate sex trafficking to larger societal problems other than crime~\citep{joh14}. Similarly, crime frames are dominant in international laws and international media outlets~\citep{aus17,gul10,gal10,sob14}. Without denying the existence of valid OR approaches focusing on criminal interdiction, framing trafficking solely from a criminal justice perspective is grossly insufficient to truly disrupt trafficking activity. As an example, there is a growing awareness that public health approaches~\citep{gre20, chi17} provide a necessary complementary viewpoint. OR and Analytics practitioners would do well to consider models that frame human trafficking interventions beyond simply criminal justice. 

\looseness-1 As evident in the review of the literature in Section \ref{Sec_Lit}, sex trafficking receives disproportionate attention from researchers and the media~\citep{aus17,den10,den16}. Content analyses of human trafficking in both United States and international media suggest that representations have predominantly focused on sex trafficking, even when broader terms such as human trafficking are invoked~\citep{sob14,den10,kin15,szo14}. Furthermore, the media commonly conflates sex trafficking with prostitution~\citep{chu14,jah05,sto07}. The overemphasis on sex trafficking in research and media is also present in prevalence studies. Some prevalence reports indicate that sex trafficking is more prevalent than labor trafficking (e.g. see the discussion in~\cite{far09}). However, this contradicts other studies that indicate labor trafficking is more prevalent (e.g., estimates from the ILO~\citep{ilo17}). These discrepancies are important to recognize. They may arise through prevalence studies that incorrectly conflate all commercial sex work with sex trafficking~\citep{far15}, as a result of sampling techniques that either lend themselves to easier identification of sex trafficking than labor trafficking (thereby introducing bias), or rely on arrest data that has been shown to disproportionally reflect sex trafficking~\citep{far14}.

\subsection{Theme 3: Not All Trafficking Activity Involves Movement} 
\looseness-1 Trafficking very often takes place without the transportation of persons across international borders. In fact, trafficking may not have anything to do with transportation or movement -- it can take place within a single country, a single community or even a single household. While there is a valid place for the growing discourse that finds similarities between trafficking and the movement of a ``product'' through a supply chain given that some instances of human trafficking do involve movement, many instances of trafficking are localized. Furthermore, analytical models that consider movement across borders must be mindful of the broader system in which trafficking occurs and the implications of enhanced policing of borders. Migrants and undocumented workers are particularly vulnerable to being trafficked, and interventions that propose to disrupt trafficking by identifying these individuals without also reforming the associated criminal justice systems that perpetuate harm to migrant and undocumented communities should be met with careful scrutiny. 

\subsection{Theme 4: Trafficking Crimes are Diverse} 
\looseness-1 Human trafficking takes diverse forms. Victims include child soldiers and child brides, domestic workers (\eb{such as }cleaners and nannies), laborers (including commercial fishing, manufacturing, construction, mining, food service, circus workers, and agricultural), sex workers (including pornography and exotic dancers), beggars and others in similarly exploitative situations.

\looseness-1 Many efforts to combat trafficking have generalized across too many types and created overly generic resources and responses~\citep{pol17}. For example, a generic interdiction model targeting to ``disrupt a trans-border trafficking network'' will be less effective than an interdiction effort tailored to explicit types of trafficking victims being trafficked across a specific border. Distinct instances of trafficking feature unique business models, trafficker profiles, recruitment strategies, victim profiles, and methods of control that facilitate human trafficking~\citep{pol17}. Individuals being trafficked across a border for escort services are often recruited with promises of modeling contracts or marriage and may encounter extreme physical and sexual violence~\citep{sur08,jon11}. \eb{While some} traffickers may be part of larger, organized networks, and in some cases may have formal or informal ties to organized crime groups such as gangs, mafia or cartels~\citep{she12,shei12} \eb{, we remind the reader that empirical evidence suggests that this is not the norm suggested by public perception}. On the other hand, individuals trafficked across borders for agricultural work are typically recruited through a more formal approach, with recruiters sometimes charging victims recruitment and travel fees that can create insurmountable debt, which becomes a control mechanism for the trafficker~\citep{bar13}.

\looseness-1 Similarly, increasing access to generalized shelter beds will be less effective than a model that considers different shelter structures that are targeted for different types of survivors. For example, while homeless shelters can provide valuable temporary protection for trafficking survivors, the lack of attention to the unique and often intensive needs of trafficking survivors can fall far short of the services that are really necessary to provide proper care and recovery. Even shelters dedicated to serving trafficking survivors must be tailored to meet the diverse needs of trafficking survivors, which among many others include child and adult survivors; survivors with undocumented legal status; LGBTQIA+ survivors; \eb{survivors with children;} and disabled survivors.  

\looseness-1 There is a misconception that trafficking roles and statuses are static. In reality, however, traffickers and victims change roles and occupy multiple statuses, both in the sex and labor trafficking contexts. For example, a victim of sex trafficking may later be ``promoted'' by the trafficker to the role of a ``bottom'' -- a victim that helps the trafficker recruit new victims and manage the trafficking operations. Similarly, an agricultural worker may move across a continuum of exploitation at a single location, ranging from labor compliance and decent work at one end of the spectrum, to extreme exploitation or forced labor at the other~\citep{skr10} (e.g. Figure~\ref{fig:Labor_trafficking _spectrum}). The interchangeability of roles, statuses and work of exploited individuals ought to be properly reflected in analytical models.

\looseness-1 Each trafficking type has unique strategies for recruiting and controlling victims and concealing the crime.  \eb{If models rooted in OR and Analytics are to succeed in assisting counter-}trafficking activities, it is imperative that the analyst understand the modus operandi of the trafficker. Polaris, a leading anti-trafficking NGO that has operated the National Trafficking Hotline for over a decade, compiled a data-driven typology of 40,000 cases of trafficking in the United States. This typology segments the market of human trafficking beyond the existing categories of sex trafficking and labor trafficking, into twenty-five distinct types of trafficking~\citep{pol17}. While the Polaris Typology is a useful starting place, as OR practitioners seeking to evaluate effective counter-trafficking responses, it is critical to forge collaborations with stakeholders in the specific industry domain to understand the nuanced differences in the characteristics of the crime across sectors.
\eb{
\subsection{Theme 5:  Trafficking Markets} 
\looseness-1 Human trafficking analysis is evolving from a narrow examination of perpetrators and victims to a more comprehensive understanding as an economic market ~\citep{aro14} in which choices are made by  traffickers, employers and consumers. Most research and practice focus on the ``supply'' side of the market: trafficked individuals. While it is critically important to understand supply-side realities and effective approaches to interdiction, victim-survivor services, and prosecution---by comparison the demand side has been largely understudied~\citep{dav14, Tay21}. 

\looseness-1 The demand-side of trafficking constitutes any individual who benefits---knowingly or unknowingly---from the exploitation of another individual. Broadly speaking, demand takes two forms: \textit{traffickers}, and \textit{consumers} whose demand drives profitability. The latter includes the end consumers of goods and services, as well as companies that consume products or services from those further up the supply chain.
Conventionally, exploiters and traffickers have been stereotypically depicted as men who are merely interested in financial profits and have no social objectives ~\citep{aro14}, and in hierarchically organised groups of criminals or loose networks of different scope  ~\citep{vog17}. Yet empirical studies show that those prosecuted as perpetrators do not seem to differ so much from their victims ~\citep{mol12,mol13,zha11}. 
A small illustrative example detailed in~\citet{vog17} describes a court case of trafficking for labor exploitation from Germany, where the trafficker and exploiter is a pregnant immigrant woman under pressure to raise money for her mother’s hospital stay. From the consumer perspective, trafficked labor is endemic and an integral part of production supply chains ~\citep{van21, gol15, bar10}. Consumers are predominately interested in a low price for goods and services.  End-consumer price sensitivity exerts severe pressure on many producers to keep costs at a minimum, often relying on illicit means with workers kept in dirty, dangerous and low-paid environments~\citep{van21}. Global supply chains are intricate, complex, and lack transparency---it is  challenging for companies, let alone consumers, to comprehend the extent of exploitation in their supply chains. Cocoa beans picked in West Africa using child labor may later be processed by forced labor, sold through an intermediary, and then produced to become candy bars. While end consumers are often unaware that the product they purchased was made by exploited labor, their demand for low-cost candy is what drives exploitative practices. 

\looseness-1 Demand-side discussions overwhelmingly focus on reducing demand for commercial sex work---not even necessarily reducing demand for sex trafficking. This has unintended negative consequences of removing income opportunities for non-trafficked sex workers that can push them into more exploitative situations.  In contrast to the focus on demand for commercial sex, discussions regarding demand for cheap goods that fuel labor trafficking is severely lacking. Curbing demand for products made with labor trafficking will require changes in policy, supply chain technologies (such as product tracking), economic analyses, public messaging, and care that increased prices for more ethically produced goods avoids unintentionally further burdening low income people. 

\looseness-1 From an OR and Analytics perspective, demand deserves consideration. Global demand---be it for sexual services or inexpensive services and products---is insatiable. A holistic approach is needed, one that not only considers the “supply-side” of the market at the expense of unintended consequences on the “demand-side”. 
}

\subsection{Theme 6: Objectives are Diverse and Nontraditional} 
\looseness-1 Working in the anti-trafficking field reveals great diversity in the organizational objectives pursued by stakeholders. For example, a nonprofit organization may have the goal of maximizing the opportunities a victim has to leave their trafficking environment, while prosecutors may have an objective of ensuring sufficient evidence is obtained for a successful prosecution. These two perspectives would likely require distinct modeling approaches. \eb{In one modeling approach, a nonprofit wishes to maximize the number of people who can safely leave their trafficking situation. In another, recognizing that traffickers replace victims who have left with new victims, the goal may be to minimize the total number that are trafficked over a time horizon} ~\citep{cau19,kos20}.

\looseness-1 This diversity in operational goals sets the objective function -- the driving force of any analytical modeling effort. While cost minimization or profit maximization is the primary motivation in many traditional OR models, anti-trafficking may necessitate other key factors as the primary objective -- including some that are difficult to quantify, such as human suffering\eb{~\citep[for instance, the notion of deprivation costs; see, e.g.,][]{hol13}}. For example, anti-trafficking efforts may be subject to budget (or other resource) constraints while minimizing harm through prevention or by helping a survivor find safe and stable housing after they leave their trafficking environment to prevent being re-trafficked. \eb{In another example, the goal may be to reduce the total number of victims over a span of time by targeting buyer disincentives and alternatives.}

\subsection{Summary of Representation of Human Trafficking} 
\looseness-1  There is great potential for OR and Analytics to successfully contribute to anti-human trafficking initiatives, and indeed as revealed in~\cite{dim21}, the nexus of OR and Analytics applied to human trafficking is emerging. To establish credibility as a field, OR and Analytics practitioners need to be conscious of the delineation of what constitutes human trafficking -- how it is defined, the diversity of crimes, and to avoid sensationalism. Not only must OR and Analytics practitioners demonstrate an understanding of \textit{what is human trafficking}, practitioners must  consider the larger ecosystem of trafficking --  vulnerable communities, respect for the individuality of victims and survivors, and the importance of multiple perspectives to provide a more holistic view of trafficking efforts. The next section discusses this larger ecosystem. 

\section{Consideration Of Survivors And Communities}\label{Sec_Communities}

We next reflect on several themes related to how OR and Analytics studies can empower survivors and communities.

\subsection{Theme 7: Value Individual Agency}
\looseness-1 Individual agency -- the notion that an individual has a fundamental human right to make a decision -- is a critical feature that OR models need to embrace.

\looseness-1 Poverty, marginalization, social and political insecurity, natural disasters and abuse increase an individual's vulnerability to exploitation and are drivers of human trafficking. Such factors push many individuals to \textit{voluntarily} seek opportunities, even via migration, to improve their well-being. The same drivers may also be a catalyst for exploitation; traffickers are known to leverage promises of basic needs, such as shelter, food, and affection to recruit and exploit individuals. Victims are lured with promises of a better life with jobs, false marriage proposals, or better lives for their children.

\looseness-1 \eb{Throughout} our work with trafficking domain experts, the concept of individual agency \eb{is} a recurring theme. This is particularly relevant to OR and Analytics interventions that propose to identify an individual or groups of individuals needing to be ``rescued''. Trafficking victims must be empowered to leave their trafficking situation when they are ready, and there are many reasons why a trafficking victim may not welcome a ``rescue'' mission or be ready to leave the trafficking circumstances at that moment. For example, such individuals may not act on this opportunity for fear of retribution, fear of law enforcement, lack of support for their children, \eb{or} lack of basic support such as shelter and food; and they may decide to stay in an exploitative situation for lack of better alternatives. Similarly, an analytical solution may identify an at-risk individual using a set of indicators at a border crossing; the individual, even after considering their individual risk tolerance, has the right to choose to cross a border and seek an opportunity to improve their livelihood if there are no other factors (e.g., legal) limiting their ability to cross\eb{~\citep[see, e.g.,][]{Dim21a}}. In a similar vein, researchers have established that incorporating a trafficking survivor's preferences into rehabilitative treatment improves outcomes~\citep{hop10,gwa19}. A survivor whose agency is not respected and who receives inadequate support after leaving their trafficking situation is at a greater risk of being re-trafficked~\citep{job10}. Thus, unless individual agency is incorporated, analytical interventions are unlikely to be successful at truly disrupting trafficking networks.

\looseness-1 Traditionally, the field of OR focuses on the development and evaluation of analytical approaches that facilitate systematic thinking and, in so doing, enables decision makers to derive viable solutions in complex settings~\citep{bec16}. Taking into consideration the concept of individual agency, there is an opportunity to rethink the normative orientation  of OR and Analytics research in the context of anti-human trafficking operations.

\subsection{Theme 8: Incorporating a Survivor-Informed Approach}
\looseness-1 Survivors offer invaluable insight and expertise as primary stakeholders in anti-HT research. Survivor engagement allows OR and Analytics researchers to incorporate nuanced aspects of human trafficking into models and proactively identify assumptions that would impede implementation or unintentionally perpetuate harm. When incorporating survivors into research efforts, the analyst needs to be mindful of their assumptions regarding survivor participation in a research group. Two illustrative examples follow. Those who have experienced being trafficked may not identify with the terms ``victim'' or ``survivor'', and they may not want their trafficking experience shared beyond the research group. Additionally, it is critical to be aware of the impact that sharing information about a traumatic, exploitative situation can have on a survivor. Be upfront with survivors regarding whether the research requires the survivor to share their own experiences, and, if so, aim to limit the number of times their experience needs to be recounted for research purposes. Furthermore, actively \emph{hire} survivors to serve on research advisory boards, and consider focusing on their general knowledge of trafficking and systems of exploitation, rather than asking detailed questions about their own individual trafficking experience. Ensure researchers have trauma-informed training and that mental health resources and referrals are readily and freely accessible for survivors involved in such research. Human trafficking survivors are an invaluable resource to OR and Analytics practitioners and intentional care must be taken to incorporate their expertise into OR and Analytics research in a way that avoids re-exploitation for their trauma narrative. Survivors should, be invited to participate in project advisory boards, guide the direction of the research, and be paid for the expertise they bring to the research.

\subsection{Theme 9: The Sum Is Greater Than The Individual Parts }
\looseness-1 The fight against HT is not the sole responsibility of a single organization or governmental body. Only a composite group of entities, each bringing their \eb{unique} skills, energy, resources and knowledge, can produce the concerted, sustained effort required to make a powerful impact in anti-trafficking initiatives. For example, modeling efforts aiming to identify trafficked individuals will likely be inadequate if shelter and corresponding rehabilitative services are not available to the survivor immediately upon leaving their trafficking situation. Analytical models suggesting solutions must be cognizant of the capacity limitations of survivor services -- a law enforcement intervention that lacks appropriate follow-up care places survivors in further jeopardy. For example, while survivors wait to access safe housing, many are unfortunately placed in a jail bed or end up back on the street, even more vulnerable to traffickers. Anti-trafficking modeling efforts have their best chance of success when accompanied by a comprehensive inclusion of diverse professionals including sociologists, criminologists, medical professionals, public health experts, law enforcement and survivors.

\subsection{Theme 10: Human Trafficking Is a Broader Community Issue}
\looseness-1 OR and Analytics researchers must be aware of the larger human trafficking ecosystem and the conditions that continue to fuel this \eb{human rights violation}. Modeling efforts are but one part of addressing human trafficking, and exploitation will continue to exist as long as there is demand. One of the most substantial causes of failure in many anti-trafficking efforts is implicit societal tolerance for exploitation. As long as people continue to exploit others for the sake of profit, the calculus of interdiction and disruption is unbalanced. A prime example is labor trafficking in apparel manufacturing. Workers in the apparel manufacturing industry, typically migrants, are often exploited and forced to work in unsafe conditions to keep production costs competitive in the global marketplace. In 2017, the value of the garment industry imported into the top G20 countries was estimated to be \$127.7 billion US, and its products are the second highest in terms of risk of being generated by trafficked labor~\citep{glo18}. The purchasing power of the G20 community increases as a benefit from the low wages and low costs of production when they purchase the apparel produced by exploitative labor conditions. Furthermore, owners and managers are not incentivized to protect and ensure the safety of workers in sectors with exploitative labor. Until labor conditions are more widely known by, and a concern to, the public, the consumer is not likely to demand the types of changes required to reverse these inequities.
Supply of vulnerable individuals is virtually unlimited; as long as racism, poverty, war, and gender-based violence exist, humans are vulnerable to exploitation and trafficking.

We mention the seemingly insurmountable odds of truly overcoming trafficking not to discourage OR and Analytics practitioners, but rather to caution against making claims that an analytical solution will, on its own, eradicate trafficking. These analytical disciplines have a long and successful history of creative, interdisciplinary work and offer a systematic approach to examine complex societal issues such as human trafficking. We encourage practitioners to study the ways in which human trafficking is dependent on, and intertwined with, other systems of oppression, and to see the value in using analytical models to address those issues as well. Addressing poverty is anti-trafficking work; addressing homelessness is anti-trafficking work; addressing racism is anti-trafficking work.

\looseness-1 In summary,  how OR and Analytics researchers approach the issue of human trafficking matters. The first five themes we presented above stress that the representation of human trafficking matters, not only for model fidelity, but for meaningful and impactful results.  Approaching any modeling efforts with the last four themes in mind illustrates to the anti-trafficking community that the  OR and Analytics communities have situational awareness and is not attempting to offer a ``cookie-cutter'' approach. With these thoughts in mind, below we suggest ways forward. 

\section{Opportunities \eb{for} OR \eb{and} Analytics Researchers}\label{Sec_Future_Research}
\looseness-1 There is a general trend for researchers to simply adapt existing methodologies developed for commercial or military operations to anti-human trafficking operations. Such efforts have resulted in a growing discourse on the topic and have been very valuable in increasing the basic knowledge of OR and Analytics to human trafficking practitioners, as well as the potential impact of working together. However, there are obvious limits to an approach of adapting existing methodologies, specifically if the complex nature of the social, economic and human conditions surrounding the trafficking context are left incomplete. This section examines ways forward for OR and Analytics researchers working in anti-human trafficking efforts. 

\subsection{Individual Agency}
\looseness-1 Further research is necessary to discern the integral role of individual and collective judgment in the success of proposed analytical solutions. While it is well-documented that behavioral factors are very important in operations research~\citep{bec16,br16,fra15,vil18}, it is still difficult to incorporate personality traits of players into the mathematical models due to the lack of appropriate theories~\citep{guo19}. A general framework or guideline to structure human judgment in vulnerable contexts would be particularly beneficial. Much of the relatively early work on behavior and OR focused on the identification of behavioral gaps between normative models and OR practice and implementation~\citep{whi16}. The fact that empirical work in this area has been limited provides evidence for the need for such a framework.

\subsection{Interdependent Systems}
\looseness-1 While anti-HT applications of Operations Research and Analytics have a growing presence in academic research, the extensive survey in~\cite{dim21} revealed that almost all existing studies examine independent interventions that lack a connection to the broader trafficking supply chain or system. Yet, there exists rich supply chain literature spanning other application areas that demonstrates how decisions in one area has multiple effects that ripple downstream, impacting subsequent operations. Such a perspective should also be incorporated into human trafficking OR and Analytics models and would help address \eb{potentially} unintended negative consequences. 

\looseness-1 For example, consider a model that seeks to reduce the prevalence of human trafficking within a city by evaluating multiple types of interventions. If the focus is only on how each intervention affects the prevalence of trafficking within the city, interventions that displace trafficking to communities immediately outside of the city lines will be evaluated as successful from the perspective of the model, even if total prevalence of trafficking within the broader geographic region remains static or even grows. The scope of the model is therefore critical to consider when designing analytical models and interpreting their results. \eb{Similarly, more research is needed to understand the  demand-side of trafficking---a typology of consumers and traffickers, their motivations, and the path they take to a purchase decision. Without consumers and their demand (such as for cheap goods and services under opaque supply chains), there is no market for  victims.}

\looseness-1 The complexity of a holistic modeling framework for intervening in interdependent systems requires a multi-agent framework with varying levels of autonomy among multiple players. A multi-agent framework should characterize how a player adapts their local decision and operational environment with limited or  delayed data, and whether any information is shared (e.g.~\cite{sre15}).

\subsection{Dynamic Perspective}
\looseness-1 Great opportunities exist for research to place more emphasis on dynamically changing conditions (volatility), the evolving role of different trafficking actors, and the use of new technology to support counter trafficking activity.  Trafficking roles and statuses are not static, and existing studies highlight the changing and dynamic behavior of trafficking networks~\citep{ kon17,kos20}. The shifting nature of illicit activity suggests that proposed models developed for a given situation may not remain relevant. Research is needed to establish how to develop models to account for timeliness and sensitivity, such as those that are robust to adjustments, and that are able to evaluate alternatives together with contingencies.

\looseness-1 Optimization models typically contain uncertain parameters -- robust optimization and stochastic optimization are two methodologies to address uncertainty (we refer to~\cite{ben09},~\cite{yan19},~\cite{ber11} and~\cite{pow19} for a detailed overview of robust and stochastic optimization frameworks and areas of further research). Uncertainty may be present not only in model parameters, but also the \eb{structure of the model} itself. Frequent adjustments to intervention strategies is common practice in industry, yet academic research has not explored this phenomenon extensively. With continued advancements in data science-intensive methods (e.g., AI and machine / deep learning applications in forecasting), it may be possible to include adaptive decisions or constraints for a hidden population. Decision support systems can continuously learn from data and behavior, providing greater adaptivity for stakeholders while improving accuracy.

\subsection {Curating Appropriate Data}
\looseness-1 Data and quantitative analysis have long been intertwined. While unstructured data exist related to trafficking activity, vulnerable populations and traffickers, harnessing and leveraging this data in a useful manner for anti-trafficking operations is still underexplored~\citep{fed14}. Similarly, few useable human trafficking datasets are currently available  to researchers.  Amalgamation and curation of data related to human trafficking are needed, presenting an opportunity for OR and Analytics practitioners to think creatively about how to (i) utilize existing data and (ii) fuse disparate data sources. 

\looseness-1 Trafficking-related data are routinely collected, yet rarely used for operational decision-support. In recent years, an emerging awareness within the OR and Analytics communities is the potential of using trafficking-related data for analysis beyond its intended purposes. It has been found that the majority of research efforts use online escort-website data to identify potential sex trafficking cases in wealthy countries~\citep{dim21}. We encourage practitioners to work with anti-HT agencies to explore if other types of existing, routinely collected data can suggest models related to operational decision-support. For example,~\cite{Dim21a} used intercept data in a novel manner to examine \eb{the efficiency of border station operations conducting transit monitoring.}

\looseness-1 In addition to identifying how to use existing data to support decision-making, the OR and Analytics communities can contribute to efforts in fusing human trafficking data -- initiatives much needed for interdisciplinary trafficking interventions.  Although (sensor) data fusion has an extensive history and is a relatively mature discipline~\citep{hal17}, it is necessary to reconsider traditional data fusion technologies, design and implementation methods to extend to new applications and environments.  For example, the proliferation of smart phones has created copious amounts of digital data and enabled ``participatory sensing''~\citep{hal17}, which presents opportunities for the OR and Analytics communities to establish methods to determine the trustworthiness of information. Another possible research vein is examining the combined use of internal (e.g. law enforcement reports related to trafficking) and external data (e.g. online advertisements) in a partially observable system.

\subsection{Holistic Objective Functions}
\looseness-1 The objective function and other components of analytical models should be adapted to not only emphasize the efficient and effective use of available resources, but also account for a holistic, socio-economic perspective with respect to any decisions to be made. Such an expanded emphasis requires that anti-trafficking models explicitly consider the opportunity costs of any actions. This may well lead to models that are structurally distinct from those inspired by traditional operations paradigms.

\looseness-1 Let us break free from the historical tendency of OR and Analytics that primarily focuses on financial metrics. Whereas financial objectives provide for a convenient common unit of analysis, where appropriate they can, and should be, infused with complementary notions from the broader spectrum of complex societal issues. 
Incorporating into the objective function aspects that represent fairness, equity, and/or advocacy engenders a richer and more complete perspective on which to base decisions. To the extent that such aspects can be translated to \eb{a common unit (even monetary) --} \textit{\eb{and many can}} -- fosters more inclusive and holistic objective functions and, thereby, outcomes.


\eb{
\subsection{Complementing Traditional OR and Analytics Methods}

\looseness-1 Traditional Operations Research and Analytics methods focus on quantitative analyses with rigid guarantees on performance and quality; these \textit{Hard OR} methods have historically worked well in improving efficiency  in a variety of traditional domains such as production, energy, finance, transportation, and supply chains.
When dealing with decisions that involve humans and their behaviors, however, outcomes generated by traditional methods can be insufficient, in the sense that difficult-to-model or unobserved factors and nuances may be left out of the modeling, which may greatly influence outcomes. 
Such factors exist in anti-human trafficking and related societal problems seeking to combat issues largely affecting marginalized populations. As OR and Analytics researchers working on such problems, it is imperative that we acknowledge the power imbalance between researchers and the marginalized communities we seek to serve. Instead of wielding this power to define what data are important and what modeling choices matter on our own, we must prioritize working alongside the populations we seek to serve and incorporate their perspective into the modelling process in meaningful ways. This is both an issues of social justice, as well as getting the models right to pursue effectiveness and equity, together with efficiency~\citep{joh20}. When applying OR and Analytics methods to domains where human lives are at stake, the lines of where to draw model boundaries may need to be extended to ensure inclusion and hearing of all voices.

A number of methodologies have emerged over the last five decades to tackle complex, ill-structured problems~\citep{rit73} and effectively engage with messy, confusing problems~\citep{sch87} that traditional, mathematically-based OR tools are unable to fully address~\citep{chu67,min11}.
These methods form what is referred to as the field of \textit{Soft OR}~\citep{min11} and include methods such as problem structuring methods (PSMs)~\citep{smi19,lam19, ack12}, cognitive mapping / SODA~\citep{ede88}, and the strategic choice approach (SCA)~\citep{fri12}.
Soft OR can serve to complement traditional quantitative OR analyses, and has been successfully employed in a wide variety of practical problem situations~\citep{min11}.
It has been argued that Soft OR practitioners enable OR practitioners to effectively engage with practice, which is an important driver for the health and development of Operations Research~\citep{dys21} and~\citep{ran15}.
Hence, to elevate the practical impact of OR and Analytics as applied to anti-trafficking and related societal problems, methods that combine (or advance the combination of) quantitative and qualitative approaches ought to be carefully considered ~\citep{sha21,mar22}. 

Technical disciplines such as OR and Analytics tend to be specialized and fragmented, largely stemming from the establishment of disciplinary boundaries~\citep{had08,max05}.
Such specialization stands in stark contrast to the vexing nature of societal challenges such as human trafficking.
There are ample opportunities for future research in OR and Analytics applied to human trafficking to widen the scope of research through \textit{transdisciplinary} approaches~\citep{max05, lan12, sha21,mar22}.
Transdisciplinary frameworks support the interface between social sciences and technical sciences, and have been used in societal challenges such as sustainability~\citep{gaz13} and energy research~\citep{spr14}.
Key features of a transdisciplinarity include (1) addressing socially-relevant problems rather than discovering generic facts; (2) employing evolving methodologies throughout the research; (3) Necessitating collaboration and coordination among disciplines to transcend disciplinary boundaries; and (4) Requiring participation of non-scientific stakeholders~\citep{gaz13}.
Incorporating Soft OR methods and transdisciplinarity provides an opportunity to both address societal problems such as human trafficking, as well as advance the fields of Operations Research and Analytics.

}



\section{Concluding Thoughts}\label{Sec_Conclusion}

\looseness-1 While human trafficking is not a new phenomenon, anti-trafficking efforts have only recently gained societal awareness. As such, new policies and interventions are constantly being implemented by \eb{government agencies}, law enforcement, nonprofit organizations, service and hospitality related businesses, faith-based groups, and healthcare professionals -- oftentimes without prior indication that such initiatives are effective or an efficient use of limited resources.

\looseness-1 Analytical models have the potential to dramatically improve anti-trafficking operations, particularly as general awareness of the phenomenon grows. Yet, the potential impact of OR and Analytics remains largely dormant amongst anti-trafficking practitioners. Simply applying traditional analytical approaches is insufficient, and while possibly well-meaning, are ultimately irresponsible -- the intricacies of humanity must be incorporated. While analogies between moving humans and moving products across a supply chain exist, it is an overly simplistic characterization of reality. Any proposed framework conceptualizing humans as ``product'' will be less effective if it does not acknowledge the human decision-making aspect, fundamental rights and agency of vulnerable individuals, and the socio-economic drivers creating conditions ripe for victimization. As \eb{the application of} OR and Analytics to disrupt human trafficking is an emerging field, substantial effort is required to ensure that the data, assumptions, and context are properly understood prior to developing new, application-specific, OR models.

\eb{ The OR and Analytics communities have the ability to move toward a shared vision of supporting social change by not only studying grand and complex challenges like human trafficking, but also using intentional modeling techniques and perspectives to propose specific new policies and interventions.  There are several success stories of our field impacting change for social good, including ongoing efforts to eradicate polio ~\citep{tho15}; stem the illicit drug trade ~\citep{beh00}; increase the number of paired kidney donations ~\citep{and15}, and inform COVID-19 safety protocols ~\citep{sch21,kes20} among many others. ~\citet{kap08} overviews several examples of how OR methods have been employed to support policy for the benefit of a community. It is our hope that the OR and Analytics communities progress toward a deeper commitment to responsible, impactful research that not only builds academic knowledge, but prioritizes working alongside communities, survivors, and practitioners to develop models that capture the nuances of human trafficking in a manner that is useful for developing policies and interventions that enact social change. This could include models and analysis that, among many others, would:
\begin{itemize}
\setlength{\itemsep}{0em}
\item Evaluate the effectiveness of various supply chain policies aimed at combating labor trafficking, 
\item Design effective incentives for companies to improve labor conditions,
\item Examine interventions aimed at buyers and curbing demand for exploitative practices, 
\item Illustrate how increased federal funding for social services and changes to current housing policy could reduce susceptibility to human trafficking by increasing access to basic needs, and
\item Model how federal policies that aim to reduce sex trafficking on online platforms can unintentionally lead to increased vulnerability in other non-internet based venues.
\end{itemize}
}

\looseness-1 We wrote this perspective piece in the spirit of sharing our experiences as OR and Analytics researchers working in this emerging field. We strongly advocate that to effect real, lasting change, researchers need to acknowledge the human element of the field. \eb{Only when} OR and Analytics practitioners are guided by the perspectives of those working in anti-trafficking fields, including survivors, medical professionals, public health experts, social workers, law-makers, and criminal justice professionals can we make meaningful and \eb{sustained} impact.

\section*{Acknowledgements}\label{Sec_Aknowledgement}
Special thanks to the  National Science Foundation  (Operations Engineering grants  CMMI 1841893 and CMII 1935602) for their support. \eb{Additionally, we are exceedingly grateful for the insights and feedback provided by Dr. Meredith Dank and Survivor Leader and LEAD student Jackie Mills provided.} 

\bibliographystyle{elsarticle-harv}\biboptions{authoryear}

\bibliography{HTRespRes}

\end{document}